\documentclass[pre,showpacs,floatfix,onecolumn,superscriptaddress]{revtex4}

\usepackage{graphicx}
\usepackage{subfigure}
\usepackage{amsmath}
\usepackage{amssymb}
\usepackage{latexsym}
\usepackage{multirow}
\usepackage{color}
\usepackage[hidelinks]{hyperref}
\usepackage{mathtools}
\usepackage{comment}
\usepackage{ulem}
\usepackage{ragged2e}
\usepackage{bm}

\begin{document}

\title{Steady state of overdamped particles in the non-conservative force field of a simple non-linear model of optical trap}

\author{Matthieu Mangeat}
\email{mangeat@lusi.uni-sb.de}
\affiliation{Univ. Bordeaux, CNRS, LOMA, UMR 5798, F-33400 Talence, France.}
\affiliation{Center for Biophysics \& Department for Theoretical Physics, Saarland University, D-66123 Saarbr{\"u}cken, Germany.}

\author{Thomas Gu\'erin}
\email{thomas.guerin@u-bordeaux.fr}
\affiliation{Univ. Bordeaux, CNRS, LOMA, UMR 5798, F-33400 Talence, France.}

\author{David S. Dean}
\email{david.dean@u-bordeaux.fr}
\affiliation{Univ. Bordeaux, CNRS, LOMA, UMR 5798, F-33400 Talence, France.}
\affiliation{Team MONC, INRIA Bordeaux Sud Ouest, CNRS UMR 5251, Bordeaux INP, Univ. Bordeaux, F-33400, Talence, France.}

\begin{abstract}
Optically trapped particles are often subject to  a non-conservative scattering force arising from radiation pressure. In this paper we present an exact solution for the steady state statistics of an overdamped Brownian particle subjected to a commonly used force field model for an optical trap. The model is the simplest of its kind that takes into account non-conservative forces. In particular, we present exact results for  certain marginals of the full three dimensional steady state probability distribution as well as results for the toroidal probability currents which are present in the steady state, as well as for the circulation of theses currents. Our analytical results are confirmed by numerical solution of the steady state Fokker-Planck equation. 
\end{abstract}

\maketitle

Optical traps and tweezers developed by Ashkin in the 1970s and 1980s~\cite{ashkin1970, ashkin2006} enable the trapping and the manipulation of nano-size particles. These optical set-ups have been used to measure very small forces in a wide range of systems, including living cells~\cite{ashkin1987} (e.g. virus, bacteria, proteins, and biopolymers), colloids~\cite{grier1997, lukic2007, jop2009}, dielectric and metallic nanoparticles~\cite{dienerowitz2008, gieseler2012, li2013, bateman2014, lehmuskero2015} and ultra-cold atoms~\cite{wildermuth2004}. Optical traps  also have  many applications, ranging from physics to biology. For instance,  they have allowed the characterization of  the elasticity of DNA strands~\cite{chu1991, wang1997, bustamante2000, huguet2010, wen2007, ritort2006}, the measurement of  ultraweak force intensity~\cite{moore2014, ranjit2016, liu2017} and have lead to propositions  for the detection of  gravitational waves~\cite{arvanitaki2013} and dark energy~\cite{rider2016}. They have also been useful to experimentally verify some theoretical results in out-of-equilibrium statistical mechanics, such as the fluctuation-dissipation theorem~\cite{collin2005, carberry2004, carberry2007, berut2016}, as well as in quantum mechanics~\cite{bassi2013}. In many of these cases, the system is forced  out-of equilibrium  by moving the trap center.

In the majority of the studies mentioned above, the optical trap is considered to be harmonic. However the trap range is obviously finite and the potential decreases far from the center of the laser beam and is thus generally anharmonic. Another reason for  anharmonicity comes from the radiation pressure of the laser which generates  a \textit{non-conservative} component to the  force. The importance of this non-conservative component  has recently been demonstrated~\cite{wu2009}. The presence of scattering forces means that  even static optical traps acting on Brownian particles   lead to an  out-of-equilibrium system which is not described by a Gibbs-Boltzmann probability distribution, and notably has non-zero steady state currents. When the particle motion is overdamped, both experimental and theoretical studies \cite{roichmann2008, sun2009, sun2010, simpson2010, moyses2015, demessieres2011} have shown that the current lines take the form of a torus, whose axis is the center of the laser beam. Similar toroidal currents have been uncovered  in the underdamped regime via experimental and theoretical analysis~\cite{amarouchene2019, mangeat2019}. In all these cases, analytical expressions for the stationary density and current have, to date, only been derived in a perturbative approach assuming small non-conservative forces~\cite{moyses2015,mangeat2019}. In fact, even in the simplest non-trivial  model including scattering forces, the motion in the longitudinal direction is driven by a \textit{colored and non-Gaussian} noise. This aspect is a major obstacle towards analytical non-perturbative theories which we wish to address in this paper. More generally, our study is an example of a characterization of currents in non-equilibrium steady states, an open problem of non-equilibrium statistical mechanics \cite{liver2020}.

In this article, we study the stationary state of overdamped Brownian particles in optical traps, in presence of a non-conservative force component created by the radiation pressure of the laser. In Section~\ref{secModel}, we first present the basic model for the optical trapping forces and then present the Langevin dynamics for the trapped particle in the overdamped limit. We consider the trapping potential to be composed of a  harmonic, conservative, component along with  the leading order  non-conservative force derived with the paraxial approximation. As well as being of direct relevance to the important field of optical trapping, the model studied constitutes a minimal model of a system which is non-equilibrium due to the presence of non-conservative forces. In particular the non-conservative force is non-Gaussian and this renders the analysis of the steady state more difficult as it cannot be simply characterized in terms of the first two moments. We show how the components of the trapped particle motion can be decomposed into motions subject to conservative forces and motions subject to non-conservative forces. In particular we define an effective two dimensional non-equilibrium process, the steady statistics of which can be used to determine the steady state statistics of the full model. In Section \ref{2d}, we show how the equilibrium distribution of the effective two dimensional process can be extracted from its Fokker-Planck equation. In Section \ref{3d}, we derive the marginal probability density function of the non-equilibrium component of the motion in the $z$ direction. Then we show how the steady state probability density and current for the full process in three dimensions can be obtained via perturbation theory, extending first order perturbative results on this model \cite{moyses2015, mangeat2019} to third order. Past studies of the model have analyzed the circulation of the steady state current perturbatively,  at the end of this section we show that these results are in fact exact to all orders in perturbation theory. In Section~\ref{secDiscussion} we conclude with a discussion about the results and an outlook.

\section{Experimentally motivated model}
\label{secModel}

We consider dielectric Brownian particles trapped in a single a laser beam, via a simplified version of the model expounded in  Refs.~\cite{gieseler2013, mangeat2019}. The Brownian particles are assumed to be spherical, dielectric and sufficiently small to be treated within the Rayleigh approximation where  their interaction with the laser can be considered solely in terms of  their dipole moments ${\bf p}$. In this model, a particle thus  experiences a force due to the electric field {\bf E} of the laser beam and a Lorentz force due to magnetic field {\bf B}. The total force can thus be written as ${\bf F} = ({\bf p}\cdot\nabla){\bf E} + \partial_t{\bf p} \times {\bf B}$, where ${\bf p}$ is the induced dipole moment of the particles. The electric field is assumed to be linearly polarized and is written as ${\bf E}({\bf x}) = \sqrt{I({\bf x})} \exp(-i\phi({\bf x})) {\bf e_x}$, where $I({\bf x})$ is the intensity. The force applied by the laser on the particles is then given by
\begin{equation}
{\bf F}({\bf x}) = \frac{1}{4}\alpha'\nabla I({\bf x}) + \frac{1}{2}\alpha''I({\bf x}) \nabla \phi({\bf x}), 
\end{equation}
with $\alpha'$ and $\alpha''$ the dispersive and dissipative parts of the polarizability, respectively. This force can then be decomposed into two contributions: ${\bf F} \simeq {\bf F_{\rm grad}} + {\bf F_{\rm scat}}$. The first one is conservative and is responsible for the particle trapping:
\begin{equation}
\label{eqFgrad}
{\bf F_{\rm grad}}({\bf x}) = - \nabla V({\bf x}) = - \kappa_x x {\bf e_x} - \kappa_y y {\bf e_y} - \kappa_z z {\bf e_z},
\end{equation}
where $\kappa_j$ is the spring constant in the direction ${\bf e_j}$ and is proportional to $\alpha'$. Here, anharmonic terms of the conservative force are omitted, since we focus on the effect of non-conservative ones. In the following we assume that the laser beam  is rotationally invariant in the $(x,y)$ plane by taking $\kappa_x=\kappa_y\equiv \kappa$ and $\kappa_z = \eta \kappa$. The second component of the force is a scattering  non-conservative force, due to the radiation pressure of the laser (and is in particular generated by the magnetic field): 
\begin{equation}
\label{eqFscat}
{\bf F_{\rm scat}}({\bf x}) = -\varepsilon \kappa a \left( 1- \frac{x^2}{a^2}-\frac{y^2}{a^2}\right) {\bf e_z},
\end{equation}
where $w_x=w_y = \sqrt{2} a$ is the beam waist of the laser in the direction $x,y$ (again equal due to rotational invariance), and $\varepsilon$ is proportional to the ratio $\alpha'/\alpha''$. Note that we have taken a direction of laser propagation different to that usually used in the literature which changes the sign of the right hand side of  Eq.~(\ref{eqFscat}) with respect to that usually used \cite{gieseler2013, mangeat2019}. The expressions of these forces are given at the first order in the paraxial approximation, i.e. for a particle staying close to the optical axis ($x=0$ and $y=0$). The exact expressions of $\kappa$ and $\varepsilon$ are given in Refs.~\cite{gieseler2013, mangeat2019}, and are not necessary to the analysis performed hereinafter. In optical traps, the harmonic restoring force in the direction along which the laser propagates is generically weaker than that in the plane perpendicular to propagation and so in the experimental context one has $\eta < 1$.

Within the  optical trap, we assume that  particle's position denoted by ${\bf X} = (X,Y,Z)$ obeys the overdamped Langevin equation:
\begin{equation}
\label{langevin}
\gamma  \frac{d {\bf X}}{dt} = {\bf F_{\rm grad}}({\bf X}) + {\bf F_{\rm scat}}({\bf X}) + \sqrt{2 k_B T\gamma}{\bm \xi}(t),
\end{equation}
where $\gamma$ is the coefficient of friction, $T$ the temperature and ${\bm \xi}$ a standard Gaussian white noise with zero mean and variance $\langle \xi_i(t) \xi_j(t') \rangle = \delta_{ij} \delta(t-t')$. Note that through out this paper the variable $\xi(t)$ (with or without subscripts) will be standard Gaussian white noise. This overdamped model is valid for particles trapped in liquids or gases that are not too rarefied, when the mass $m$ of particles satisfies $m \ll \gamma^2/\kappa$.
From Eqs.~(\ref{eqFgrad}) and~(\ref{eqFscat}),  the Langevin equation gives
\begin{gather}
\dot X(t) + \lambda X(t)  = \sqrt{2D}\xi_X(t), \label{langoverx}\\
\dot Y(t) + \lambda Y(t)  = \sqrt{2D}\xi_Y(t), \label{langovery}\\
\dot Z(t) + \eta \lambda Z(t) = -\varepsilon \lambda a \left[ 1-\frac{X(t)^2+Y(t)^2}{a^2} \right] + \sqrt{2D}\xi_Z(t), \label{langoverz}
\end{gather}
with the damping rate $\lambda = \kappa/\gamma$ and the microscopic diffusion constant $D= k_B T / \gamma$, from the Einstein relation. We thus see that $X(t)$ and $Y(t)$ are then identical and independent Ornstein-Uhlenbeck (OU) processes, while $Z(t)$ is subject to a harmonic plus  a non-conservative force. 

In Eq. (\ref{langoverz}), we see that the constant term in the radiation pressure $-\varepsilon\lambda a$ can be removed by an appropriate shift of the $Z$-coordinate. Furthermore, the analysis is simplified by  writing the coordinates $X,\ Y,$ and $Z$ in terms of the widths of the harmonic oscillator (in absence of non-conservative force), we thus write:
\begin{equation}
X= \sqrt{\frac{k_BT}{\kappa}} x,\quad Y= \sqrt{\frac{k_BT}{\kappa}} y, \quad Z= -\frac{\varepsilon a}{\eta}+\sqrt{\frac{k_BT}{\eta\kappa}} z.
\end{equation}
We also work in terms of the rescaled time $\tau =\lambda t$ and  these rescalings lead to  
\begin{gather}
\dot x(\tau) + x(\tau)  = \sqrt{2}\xi_x(\tau), \label{langoverX}\\
\dot y(\tau) + y(\tau)  = \sqrt{2}\xi_y(\tau), \label{langoverY}\\
\dot z(\tau) + \eta z(\tau) = \frac{\eta^2\epsilon}{4}  \left[ x(\tau)^2+y(\tau)^2 \right]  + \sqrt{2\eta }\xi_z(\tau),\label{langoverZ}
\end{gather}
where the intensity of the scattering force in these rescaled variables is given by
\begin{align}
\epsilon= \frac{4\varepsilon}{a \eta} \sqrt{\frac{k_BT}{\eta\kappa}}.
\end{align}

We consider the probability density $p(x,y,z,\tau)$ to observe $x,y,z$ at time $\tau$, this density obeys the Fokker-Planck equation:
\begin{align}
\frac{\partial p}{\partial \tau}(x,y,z;\tau) &= \frac{\partial}{\partial x}\left[\frac{\partial p}{\partial x}(x,y,z;\tau) + x p(x,y,z;\tau)\right] + \frac{\partial}{\partial y}\left[\frac{\partial p}{\partial y}(x,y,z;\tau) + y p(x,y,z;\tau)\right]\nonumber \\
&+ \eta\frac{\partial}{\partial z} \left\{ \frac{\partial p}{\partial z}(x,y,z;\tau)+\left[z-\frac{\eta\epsilon}{4}(x^2+y^2) \right]p(x,y,z;\tau)\right\}.\label{ffp}
\end{align}
Due to the linearity of the equation for $z(\tau)$, Eq.~(\ref{langoverZ}), we decompose the motion in the $z$ direction into a damped harmonic oscillator type motion plus a term generated by the non-conservative forces: $z(\tau) = z_{o}(\tau) + z_{n}(\tau)$. The OU component $z_o(\tau)$ then obeys
\begin{equation}
\dot z_{o}(\tau) + \eta z_{o}(\tau) =   \sqrt{2\eta}\xi_z(\tau),
\end{equation}
and the term $z_n(\tau)$ generated by the non-conservative forces thus obeys
\begin{equation}
\dot z_n(\tau) + \eta z_n(\tau) = \frac{\eta^2\epsilon}{4}  [x(\tau)^2+y(\tau)^2].
\end{equation}
This equation can be formally integrated, assuming that $z_n(0)=0$ which has no effect on the late time equilibrium distribution we find that
\begin{equation}
 z_n(\tau)=\frac{\eta^2\epsilon}{4} \int_0^\tau d\tau' \exp[-\eta(\tau-\tau')] [x(\tau')^2+y(\tau')^2] \label{rept}.
\end{equation}
This above representation can be used to derive the equilibrium statistics of $z_n(\tau)$ by evaluating its generating function within a path integral formulation. We will not use this method in this paper however as the main result can be derived directly from the corresponding Fokker-Planck equation. 

The process $z_n$ can in turn be decomposed as 
$z_n(\tau)= z_{nx}(\tau)+z_{ny}(\tau)$ where 
\begin{eqnarray} 
\dot z_{nx}(\tau) +\eta z_{nx}(\tau) &=&\frac{\eta^2\epsilon}{4} x(\tau)^2 \\
\dot z_{ny}(\tau) +\eta  z_{ny}(\tau) &=&\frac{\eta^2\epsilon}{4} y(\tau)^2,
\end{eqnarray}
and given the independence, and statistical equivalence, of $x(t)$ and $y(t)$ we see that $z_{nx}$ and $z_{ny}$ are  independent and identically distributed. 
This means we can write the following identities in law of stochastic processes
\begin{equation}
z_{nx}(\tau)\equiv z_{ny}(\tau)\equiv \chi(\tau),
\end{equation}
where $\chi(\tau)$ is driven by the OU process
\begin{equation}
\dot x(\tau) +  x(\tau)  = \sqrt{2}\xi(\tau),\label{xn}
\end{equation}
via the equation
\begin{equation}
\dot \chi(\tau) +\eta\chi(\tau) =\frac{\eta^2\epsilon }{4} x(\tau)^2.\label{zeta}
\end{equation}
With $\chi(\tau)$ defined as above, we see that $\chi(\tau)$  must be positive for all times once it crosses $\chi=0$ for the first time due to the driving term being positive. This means that the equilibrium distribution must have as support positive values of $\chi$. The full statistics of the three dimensional model can thus be deduced from the effective two dimensional model given by Eq.~(\ref{xn}) and Eq.~(\ref{zeta}). While Eq.~(\ref{zeta}) appears relatively simple, the fact that the driving noise $x^2(\tau)$ is non-Gaussian and colored significantly complicates the analysis. 

\section{Equilibrium statistics of the effective two dimensional model}
\label{2d}

Here we analyze Eq.~(\ref{xn}) and Eq.~(\ref{zeta}) which in themselves represent a minimal model for a stochastic process driven by non-conservative forces and is thus interesting in its own right. The Fokker-Planck equation for the pair $(x(\tau),\chi(\tau))$ is  given by
\begin{equation}
\frac{\partial p}{\partial \tau}(x,\chi;\tau) = \frac{\partial}{\partial x}\left[\frac{\partial p}{\partial x}(x,\chi;\tau) + xp(x,\chi;\tau)\right]-\eta \frac{\partial}{\partial \chi}\left[\left(\frac{\epsilon\eta}{4} x^2 - \chi\right)p(x,\chi;\tau)\right],
\end{equation}
and the steady state distribution $p_{s}(x,\chi)= p(x,\chi;\tau\to\infty)$ we are interested in   satisfies
\begin{equation}
\frac{\partial}{\partial x}\left[\frac{\partial p_{s}}{\partial x}(x,\chi)+xp_{s}(x,\chi)\right]-\eta \frac{\partial}{\partial \chi}\left[ \left(\frac{\epsilon\eta}{4} x^2 - \chi\right) p_{s}(x,\chi)\right]=0.
\end{equation}
 Noting that $p_s(x,\chi)$ vanishes for negative $\chi$, we take the Laplace transform of $p_{s}(x,\chi)$ with respect to $\chi$,
\begin{equation}
\tilde p_{s}(x,s)= \int_0^\infty d\chi\  p_{s}(x,\chi)\exp(-s \chi),
\end{equation}
we see that this obeys
\begin{equation}
\frac{\partial}{\partial x}\left[\frac{\partial \tilde p_{s}}{\partial x}(x,s)+x\tilde p_{s}(x,s)\right]-\eta s\left[ \frac{\epsilon\eta }{4} x^2 \tilde p_{s}(x,s)+ \frac{\partial \tilde p_{s}}{\partial s}(x,s)\right]=0,
\end{equation}
where we assumed $p_s(x,\chi=0)=0$, which must be the case due to the fact that  of the driving term in Eq. (\ref{zeta}) is always positive. 

We now analyze the problem by adapting the method used in Ref. \cite{boyer2011} to study optimal fitting schemes for estimators of the diffusion constant of Brownian motion. We look for a solution of the form
\begin{equation}
\tilde p_{s}(x,s)= B(s)\exp\left[-\frac{1}{2}A(s)x^2\right]. \label{Eqps2D}
\end{equation}
The fact that this ansatz works is intrinsically related to the quadratic dependence of $\chi$ on $x$ and $y$ in the representation given in Eq.~(\ref{rept}).
This yields the coupled ordinary differential equations
\begin{gather}
s \eta A'(s) = 2 A(s)(1-A(s))+ \frac{s\epsilon\eta^2}{2},\label{eA}\\
s\eta B'(s) = B(s)(1 -  A(s)).\label{eB}
\end{gather}
When $s=0$ we must recover the   marginal distribution of the OU process $x(\tau)$, this  tells us that
\begin{equation}
\tilde p_{s}(x,0) = \frac{1}{\sqrt{2\pi }}\exp\left(-\frac{x^2}{2}\right),
\end{equation}
which then gives
\begin{equation}
A(0) = 1 \quad {\rm and} \quad B(0) = \frac{1}{\sqrt{2\pi}}.\label{bcab}
\end{equation}
The equation for $A$ can be integrated via the ansatz $A(s)=g s h'(s)/h(s)$ which on choosing
$g=\eta/2$ renders the resulting differential equation linear:
\begin{equation}
s h''(s)  + (1-\alpha) h'(s) - \epsilon  h(s)=0,
\end{equation}
with  $\alpha = 2/\eta$. The linearly independent solutions to this equation are 
\begin{equation}
h(s) = s^{\frac{\alpha}{2}}I_{\alpha}(2\sqrt{s\epsilon}) \quad {\rm and} \quad h(s) =  s^{\frac{\alpha}{2}}I_{-\alpha}(2\sqrt{s\epsilon}),
\end{equation}
where $I_\nu(y)$ is the modified Bessel function of the first kind with order $\nu$ \cite{abram}.
The small argument behavior of the modified Bessel function is
\begin{equation}
I_\nu(y) \simeq \frac{(y/2)^\nu}{\Gamma(1+\nu)},
\end{equation}
where $\Gamma(y)$ is the Gamma function. Using the boundary condition for $A(s)$ in Eq.~(\ref{bcab}) we find
that 
\begin{equation}
h(s) = s^{\frac{\alpha}{2}}I_{\alpha}(2\sqrt{s\epsilon}),
\end{equation}
up to an overall unimportant constant. From this we find that $B(s)$ obeys
\begin{equation}
\frac{B'(s)}{B(s)} = \frac{\alpha}{2s} -\frac{1}{2}\frac{h'(s)}{h(s)},
\end{equation}
and then
\begin{equation}
B(s) = \frac{(s\epsilon)^{\frac{\alpha}{4}}}{\sqrt{2\pi \alpha\Gamma(\alpha)  I_\alpha(2\sqrt{s\epsilon})}}.\label{EqB}
\end{equation}
We also find that $A(s)$ is given explicitly by
\begin{equation}
A(s) = \frac{\sqrt{s\epsilon }}{\alpha}\frac{I_{\alpha-1}(2\sqrt{s\epsilon})}{I_{\alpha}(2\sqrt{s\epsilon})}.\label{EqA}
\end{equation}
The function $A(s)$ turns out to be identical, up to a rescaling of $s$, to the Laplace transform (with respect to time) of the rate of creep function of so called Bessel viscoelastic models \cite{colo2017}.
In summary, in this Section, we have derived an explicit expression  [Eqs.~(\ref{Eqps2D}), (\ref{EqB}) and (\ref{EqA})] in Laplace space for the joint probability distribution of $(x,\chi)$ for the effective two-dimensional model. These results will prove useful for the analysis of the three-dimensional trap. 

\section{Probability distribution for   the three dimensional trap}\label{3d}

Here we use the results of the previous section to analyze the steady state in the three dimensional trap. To start with we analyze the distribution of $(x,y,z_n)$, in particular concentrating on the marginal distribution of $z_n$, and then consider the distribution of $(x,y,z)$. Note that studying the distribution of $(x,y,z_n)$ only can be viewed as the limit $\varepsilon=\infty$ of the model, since in this case $z$ is driven by scattering forces with negligible thermal noise.

\subsection{The distribution of the non-equilibrium component of the displacement}
Returning to the full three dimensional problem we find that the Laplace transform of the stationary probability distribution for the variables $x,\ y$ and $z_n$, denoted by $p_{ns}(x,y,z_n)$, given by
\begin{equation}
\tilde p_{ns}(x,y,s)=\int_0^\infty dz_n\ p_{ns}(x,y,z_n)\exp(-sz_n),
\end{equation}
is given by
\begin{equation}
\tilde p_{ns}(x,y,s)= B^2(s)\exp\left[-\frac{A(s)}{2}(x^2+y^2)\right]=\frac{(s\epsilon)^{\frac{\alpha}{2}}}{2\pi \alpha\Gamma(\alpha)  I_\alpha(2\sqrt{s\epsilon})}\exp\left[-\frac{\sqrt{s\epsilon }}{2\alpha}\frac{I_{\alpha-1}(2\sqrt{s\epsilon})}{I_{\alpha}(2\sqrt{s\epsilon})}(x^2+y^2)\right].\label{fulln}
\end{equation}
The Laplace transform of the marginal probability density function of the displacement $z_n$ due to the non-conservative force  $\tilde p_{ns}(s)$ is given by
\begin{equation}
\tilde p_{ns}(s)= \int dx dy \  \tilde p_{ns}(x,y,s)= \frac{(\epsilon s)^{\frac{\alpha-1}{2}}}{ \Gamma(\alpha)  I_{\alpha-1}(2\sqrt{\epsilon s})}.\label{pm}
\end{equation}
Remarkably Eq.~(\ref{pm}) shows that the equilibrium distribution of $z_n$ is the same as  that of the first hitting time at $\sqrt{2\epsilon}$ of a Bessel process of index $\nu= \alpha-1$ started at $0$ \cite{kent1978,geto1979}. The Bessel process of order $\nu$ can be interpreted as the radial part of a Brownian motion in $d= 2\nu+2$ dimensions, so here we have $d=2\alpha$. This observation may simply be related to the appearance of Bessel's equation in our analysis but nonetheless is rather intriguing given the seemingly very different nature of the two stochastic processes involved. 

\begin{figure}[t]
\begin{center}
  \includegraphics[width=16cm]{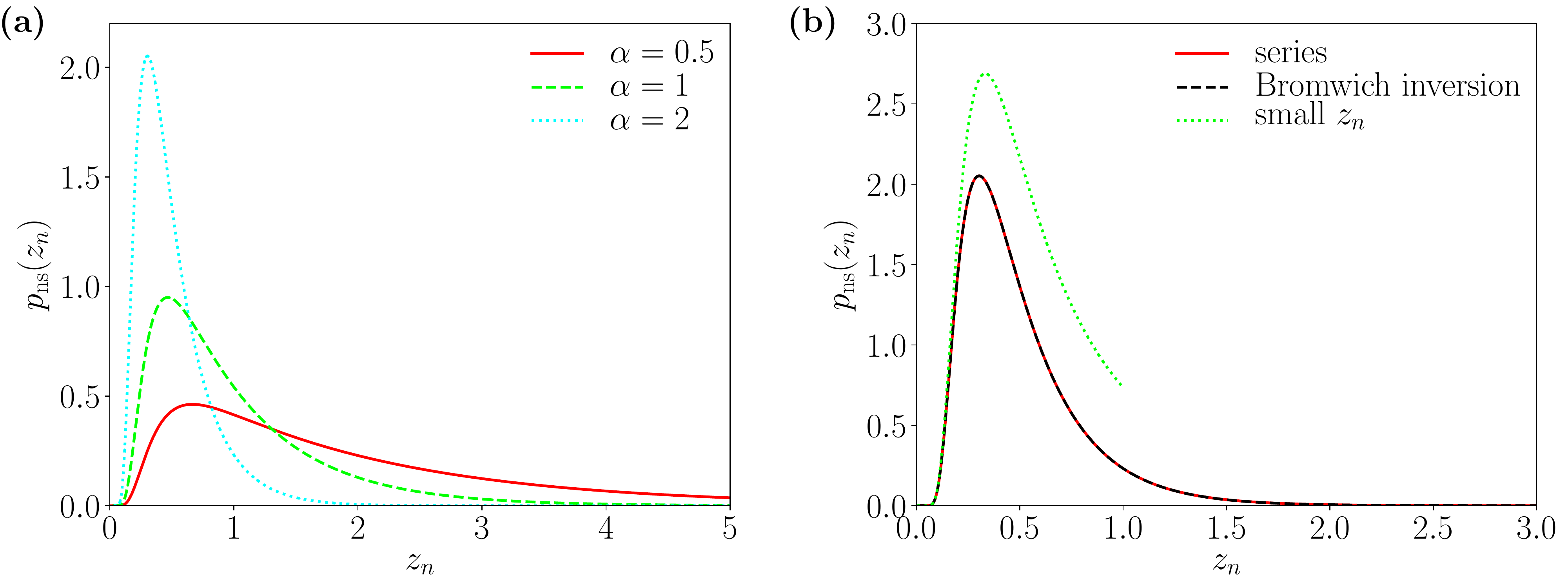}
  \caption{{\bf (a)}~The function $p_{ns}(z_n)$ for evaluated using the representation Eq.~(\ref{psex}) by taking into account the zeros of  $J_{\alpha-1}(u)$ on $(0,1000]$. Shown are the function for $\alpha=1/2$ (solid line),
  $\alpha=1$ (dashed line) and $\alpha =2$ (dotted line). {\bf (b)}~The function $p_{ns}(z_n)$ for $\alpha=2$, evaluated using the representation Eq.~(\ref{psex}) by taking into account the zeros of  $J_{\alpha-1}(u)$ on $(0,1000]$ (solid line) compared with numerical computation of the Bromwich inversion formula Eq.~(\ref{Itin}) (dashed line), the two curves are indistinguishable. Also shown is the analytic prediction for the behavior at small $z_n$ given by Eq.~(\ref{ess}) (dotted line). Note that $\epsilon=1$, without loss of generality.}  \label{figpns}\label{figcompar}
\end{center}
\end{figure}

The cumulant generating function for $z_n$ is given by 
\begin{equation}
M(s)= \sum_{k=1}^\infty \frac{(-1)^k}{k!}\langle z_n^n\rangle_c \ s^k= \ln(\tilde p_{ns}(s)),
\end{equation}
where $\langle z^n\rangle_c$ denotes the n$^{\rm th}$ cumulant, or equivalently the connected part of
the  n$^{\rm th}$ moment. Using  the series expansion of Bessel functions in Eq.~(\ref{pm}), we find
\begin{eqnarray}
M(s) = -\ln\left(\sum_{k=0}^\infty \frac{\Gamma(\alpha)}{k!\ \Gamma(k+\alpha)}\epsilon^ks^{k}\right),
\end{eqnarray}
From this we obtain the first four cumulants as
\begin{align}
\langle z_n\rangle_c&= \frac{\epsilon}{\alpha},\nonumber \\
\langle z_n^2\rangle_c&= \frac{\epsilon^2}{\alpha^2(1+\alpha)},\nonumber \\
\langle z_n^3\rangle_c&= \frac{4\epsilon^3}{\alpha^3(\alpha+1)(\alpha+2)},\nonumber \\
\langle z_n^4\rangle_c&= \frac{6\epsilon^4(6+5\alpha)}{\alpha^4(1+\alpha)^2(\alpha+2)(\alpha+3)}.
\end{align}
Furthermore, as  $\tilde p_{ns}(s)$ has an expansion in terms of integer powers of $s$, we see that the moments exist at all orders and moreover we see that the function $\tilde p_{ns}(s)$ is single valued in the complex plane. With this in mind, we use the Bromwich inversion formula to write the density of $z_n$ as
\begin{equation}
p_{ns}(z_n) =\frac{1}{\epsilon\Gamma(\alpha)} \int_\gamma \frac{ds}{2\pi i} \frac{ s^{\frac{\alpha-1}{2}}\exp(s\frac{z_n}{\epsilon})}{ I_{\alpha-1}(2\sqrt{s})}.\label{Itin}
\end{equation}
The poles of the integrand lie along the negative real axis and the Bromwich inversion contour   $\gamma$  lies to the right of these poles. When $z_n<0$ this means that $p_{ns}(z_n)=0$, and for 
$z_n>0$ standard complex analysis shows that
\begin{equation}
p_{ns}(z_n) = \frac{1}{\epsilon\Gamma(\alpha)}\sum_{k=1}^\infty \left(\frac{u_{\alpha k}}{2}\right)^{\alpha}\frac{\exp(-\frac{u^2_{\alpha k}}{4\epsilon}z_n)}{J_\alpha(u_{\alpha k})},\label{psex}
\end{equation}
where $J_\nu$ is the modified Bessel function of the second kind and is related to $I_\nu$ via $I_\nu(y)=i^{-\nu}J_\nu(iy)$. The terms $u_{\alpha k}$ correspond to the $k^{\rm th}$ positive, non zero, root of the equation $J_{\alpha-1}(u)=0$. From this we can immediately see that for large $z_n$ one has
\begin{equation}
p_{ns}(z_n) \simeq \frac{1}{\epsilon \Gamma(\alpha)}\left(\frac{u_{\alpha 1}}{2}\right)^{\alpha}\frac{\exp(-\frac{u^2_{\alpha 1}}{4\epsilon}z_n)}{J_\alpha(u_{\alpha 1})}.
\end{equation}
The small $z_n$ behavior of $p_{ns}(z_n)$ can be extracted from the large $s$ asymptotic behavior of $\tilde p_{ns}(s)$. Using the asymptotic form for the modified Bessel function
\begin{equation}
I_\nu(z) \underset{z\to\infty}{\simeq} \frac{\exp(z)}{\sqrt{2\pi z}},  
\end{equation}
we find
\begin{equation}
\tilde p_{ns}(s/\epsilon)\simeq 2\sqrt{\pi} \frac{s^{\frac{2\alpha-1}{4}}\exp(-2\sqrt{s})}{\Gamma(\alpha)}.\label{pnas}
\end{equation}
Inserting this form into Eq.~(\ref{Itin}) and making the change of variables $s=\epsilon^2u/z_n^2$ the integral can be evaluated with the saddle-point method, with $1/z_n$ the large parameter, to obtain
\begin{equation}
p_{ns}(z_n) \simeq \frac{2\epsilon^\alpha}{\Gamma(\alpha)z_n^{\alpha+1}}\exp\left(-\frac{\epsilon}{z_n}\right).\label{ess}
\end{equation}
To derive the above one may also use the of  method of inspection used in \cite{boyer2011} where a similar asymptotic behavior arises from the consideration of a quadratic path integral arising in the analysis of fitting procedures to evaluate the diffusion constant of Brownian motion.
While in the above we have maintained the dependence on $\epsilon$ to examine the full problem, it is obvious from the beginning that $z_n = \epsilon\zeta_n$ where $\zeta_n$ is independent of $\epsilon$. We can thus consider, without loss of generality, the case where $\epsilon=1$. 
The sum given in Eq.~(\ref{psex}) can be evaluated numerically and as long as $z_n$ is not too close to $0$ a finite number of terms give a good approximation to the full result. In Fig.~\ref{figpns}(a) we show the form of $p_{ns}(z_n)$ as a function of $z_n$ where we have taken all the positive  zeros of the $J_{\alpha-1}(u)$ on $(0,1000]$. We  see that these forms of $p_{ns}$ exhibit, as predicted above an exponential decay for large $z_n$ and one sees the rapid decay to zero as $z_n\to 0$, however more and more terms are required to reproduce the essential singularity at the origin. The distributions are shown for $\alpha=1/2$, $\alpha=1$ and $\alpha=2$.
One can check that the numerical form of these distributions away from $z_n=0$ does not change on extending the interval over which the zeros of $J_{\alpha-1}(u)$ are taken.

The inverse Laplace transform in Eq.~(\ref{Itin}) can also be evaluated numerically, for instance using the numerical integration option of Mathematica, this turns out to be relatively simple to do and in Fig.~\ref{figcompar}(b) we show the results for $\alpha=2$ from the infinite series representation Eq.~(\ref{psex}) (again using the zeros of $J_{\alpha-1}(u)$ on $(0,1000]$) compared with the numerical inversion of the Laplace transform. We see that the agreement is perfect, however in both cases numerical precision becomes an issue for small $z_n$ due to the presence of the essential singularity exhibited in Eq.~(\ref{ess}). The asymptotic expression for small $z_n$ Eq.~(\ref{ess}) is also shown, we see that while it predicts the position of the maximum of the probability density function it over estimates its height. Of course, on approaching $z_n=0$ the agreement is perfect.

The joint probability distribution for $(x,y,z_n)$ is more difficult to analyze. However we may again examine the form of the distribution for small $z_n$ using the asymptotic behavior given in Eq.~(\ref{fulln}) for large $s$ which leads to
\begin{equation}
p_{ns}(x,y,z_n)= \frac{\epsilon^{\alpha+1}}{\pi\alpha\Gamma(\alpha) z_n^{\alpha+2}}\left(1+\frac{\rho^2}{4\alpha}\right)^{\alpha+\frac{3}{2}}\exp\left[-\frac{\epsilon}{z_n}\left(1+\frac{\rho^2}{4\alpha}\right)^2\right],\label{smallz}
\end{equation}
where $\rho = \sqrt{x^2+y^2}$.

The steady state current corresponding to the processes $x,y$ and $z_n$ is given by
\begin{gather}
{\bf J}_{ns}(x,y,z_n) = -\left[\frac{\partial p_{ns}}{\partial x}(x,y,z_n) + xp_{ns}(x,y,z_n)\right]{\bf e}_x -\left[\frac{\partial p_{ns}}{\partial y}(x,y,z_n) + y p_{ns}(x,y,z_n)\right]{\bf e}_y \nonumber \\
+\eta\left[\left(\frac{\epsilon\eta}{4}(x^2+y^2)-z_n \right)p_{ns}(x,y,z_n)\right]{\bf e}_z,\label{jns}
\end{gather}
and using the rotational symmetry of the problem we can write this as
\begin{equation}
{\bf J}_{ns}(\rho,z_n) = p_{ns}(\rho,z_n)\left[-\left(\frac{\partial\ln p_{ns}}{\partial \rho}(\rho,z_n) + \rho \right){\bf e}_\rho +\frac{2}{\alpha}\left(\frac{\epsilon}{2\alpha}\rho^2-z_n \right){\bf e}_z\right]
\end{equation}
where ${\bf e}_\rho = (x{\bf e}_x + y{\bf e}_y)/\rho$ is the polar basis vector and we recall that $\eta = 2/\alpha$. In the region where $z_n\ll 1$ we can then use the asymptotic result Eq.~(\ref{smallz}) to find
\begin{eqnarray}
{\bf J}_{ns}(\rho,z_n)&\simeq& \frac{\rho\epsilon^{\alpha+2}}{\pi\alpha^2\Gamma(\alpha) z_n^{\alpha+2}}\left(1+\frac{\rho^2}{4\alpha}\right)^{\alpha+\frac{3}{2}}\exp\left[-\frac{1}{z_n}\left(1+\frac{\rho^2}{4\alpha}\right)^2\right]
\left[ \frac{1}{z_n }\left(1+\frac{\rho^2}{4\alpha}\right){\bf e}_\rho +\frac{1}{\alpha}\rho{\bf e}_z\right].
\end{eqnarray}
We will see in what follows that the full steady state distribution for the process $(x,y,z)=(x,y,z_n+z_o)$ as well as the corresponding current is simply related to the corresponding results for $(x,y,z_n)$. However as 
we cannot obtain a fully analytical form for the distribution of the latter we must resort to a perturbative analysis. 

\subsection{Numerical results for the process $(x,y,z_n)$}

\begin{figure}[t]
\begin{center}
\includegraphics[width=16cm]{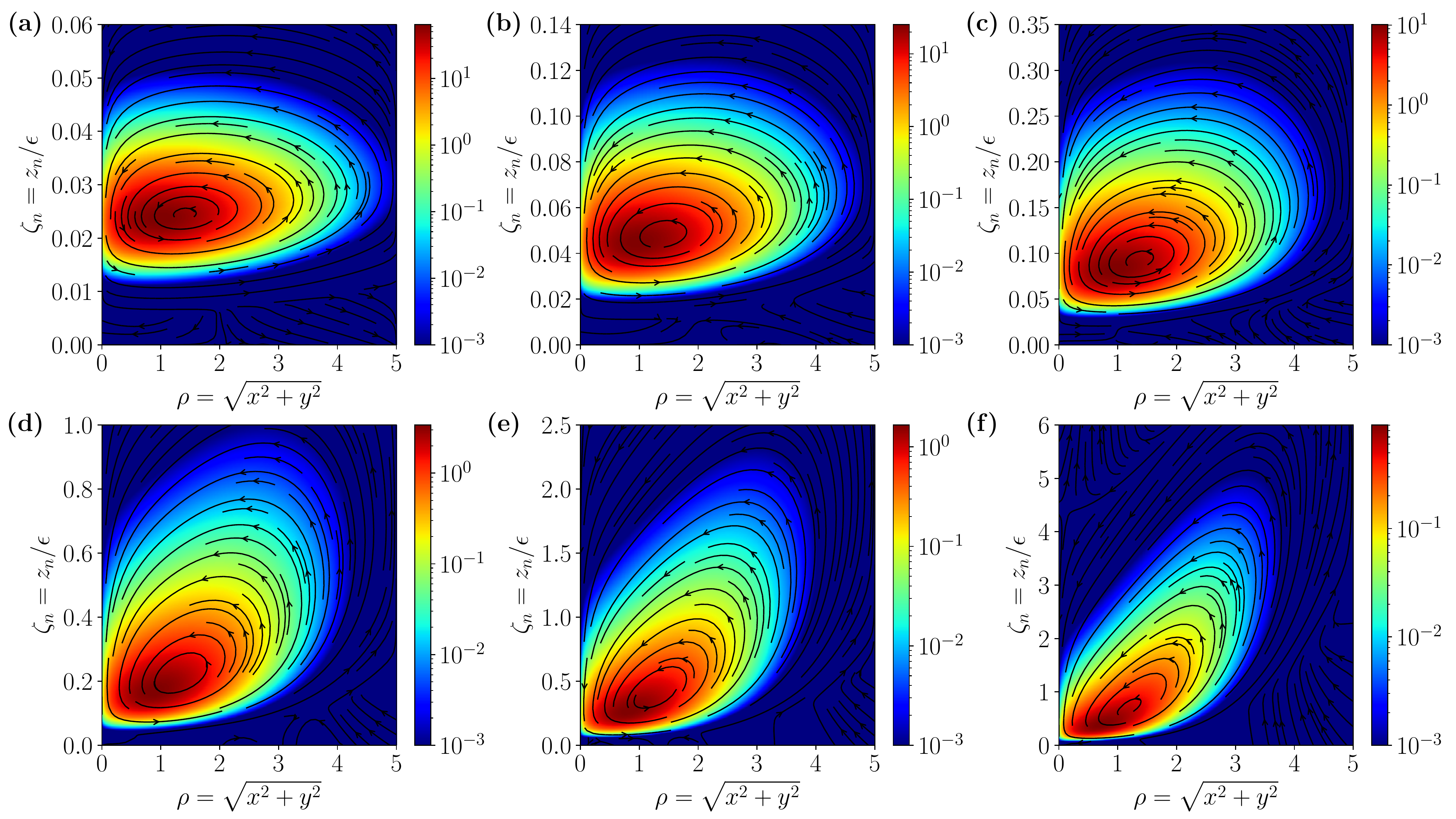}
\end{center}
\caption{The probability distribution $p_{cns1}(\rho,\zeta_n)$ for the process $(\rho,\zeta_n)$ along with the associated current lines for various values of $\eta=2/\alpha$: {\bf (a)} $\eta=0.05$, {\bf (b)} $\eta=0.1$, {\bf (c)} $\eta=0.2$, {\bf (d)} $\eta=0.5$, {\bf (e)} $\eta=1$, and {\bf (f)} $\eta=2$. In all cases there is a single vortex. These results were obtained by integrating numerically the steady state Fokker-Planck equation for $(\rho,\zeta_n)$ with the finite element method.} \label{nsvort}
\end{figure}

The most straightforward way to determine the steady state distribution of the process $(x,y,z_n)$ is to solve the time independent Fokker-Planck equation for the process in the case where we set $\epsilon=1$
to numerically evaluate what we denote by $p_{ns1}(x,y,z_n)$. The distribution for general $\epsilon$ is then simply obtained via 
\begin{equation}
p_{ns}(x,y,z_n)= \epsilon p_{ns1}(x,y,\epsilon z_n).
\end{equation}
We thus solve the steady state diffusion equation using the current given by Eq.~(\ref{jns}). Given the essential singularity at $z_n=0$ the numerical resolution is difficult in this region, however this problem can be surmounted by adding a small diffusive component to the process $z_n$. In our computations a diffusion constant of  $10^{-6}$ was taken. This is equivalent to numerically solving the steady state  Fokker-Planck equation for $\epsilon=1000$ and then extracting the probability density function for $\zeta_n=z_n/\epsilon$. In this analysis the only parameter corresponds to $\alpha = 2/\eta$. Shown in Fig.~\ref{nsvort} is the stationary probability density for the cylindrical coordinate process $(\rho,\zeta_n)$, denoted by  $p_{cns1}$ and which is simply related to $p_{ns1}$ via
\begin{equation}
p_{cns1}(\rho,\zeta_n) =2\pi \rho p_{ns1}(\rho, \zeta_n),\label{cartoc}
\end{equation}
for several values of $\eta$. On the same figure we also  the associated current as a function of $\zeta_n$ and the radial coordinate $\rho$.  We see that in all cases there is a single current vortex which is located close to the maximum of the probability density function $p_{cns1}(\rho,\zeta_n)$. 

\subsection{Distribution for the full model}

In the previous analysis we did  not include the OU contribution to the movement in the $z$ direction. If we denote  full distribution $p_{fs}(x,y,z)$ we see that its two sided Laplace transform with respect to $z$ is given by
\begin{equation}
\tilde p_{fs}(x,y,s)= \tilde p_{ns}(x,y,s)\exp\left(\frac{s^2}{2}\right)=\frac{(s\epsilon)^{\frac{\alpha}{2}}}{2\pi \alpha\Gamma(\alpha) I_\alpha(2\sqrt{s\epsilon})}\exp\left(\frac{s^2}{2}\right)\exp\left[-\frac{\sqrt{s\epsilon }}{2\alpha}\frac{I_{\alpha-1}(2\sqrt{s\epsilon})}{I_{\alpha}(2\sqrt{s\epsilon})}(x^2+y^2)\right].\label{pfsl}
\end{equation}
Note that we take the two sided Laplace transform as the total process $z$ is can now become negative.

The full marginal distribution for  $z$ can obviously be derived from the convolution with the pdf of the OU component of the $z$ motion, that is to say
\begin{equation}
p_{fs}(z) = \frac{1}{\sqrt{2\pi}}\int_0^\infty dz_n \exp\left[-\frac{1}{2}(z-z_n)^2\right]p_{ns}(z_n) .
\end{equation}

However, to  make the link with previous studies at the perturbative level, of  the effect of the non-conservative force on the steady state distribution we now proceed by expanding  Eq.~(\ref{pfsl}) in $\epsilon$. We find that the form of this expansion is 
\begin{equation}
\tilde p_{fs}(\rho,s)= \frac{1}{2\pi} \exp\left(\frac{s^2-\rho^2}{2}\right) \sum_{k=0}^\infty\epsilon^k s^k C_k(\rho,\alpha),\label{0431}
\end{equation}
where we again emphasize that $p_{fs}$ is still the probability density function for $(x,y,z)$ but we have written it as a function of $(\rho,z)$ for compactness. 
Carrying out the above expansion explicitly, we find the first three terms:
\begin{align}
C_0(\rho,\alpha) &= 1 \\
C_1(\rho,\alpha) &= - \frac{2\alpha + \rho^2}{2\alpha(\alpha+1)} \\
C_2(\rho,\alpha) &= \frac{4\alpha^2(3+\alpha)+ 4\alpha(3+\alpha)\rho^2+(2+\alpha) \rho^4}{8\alpha^2(1+\alpha)^2(2+\alpha)}\\
C_3(\rho,\alpha) &= -\frac{(8 \alpha ^3+12\alpha^2 \rho^2) (\alpha^2+ 8 \alpha +19)+6 \alpha  (\alpha +3) (\alpha +4) \rho ^4+(\alpha +2) (\alpha +3) \rho ^6}{48 \alpha ^3 (\alpha +1)^3 (\alpha +2) (\alpha +3)}.
\end{align}

\begin{figure}[t]
\begin{center}
\includegraphics[width=16cm]{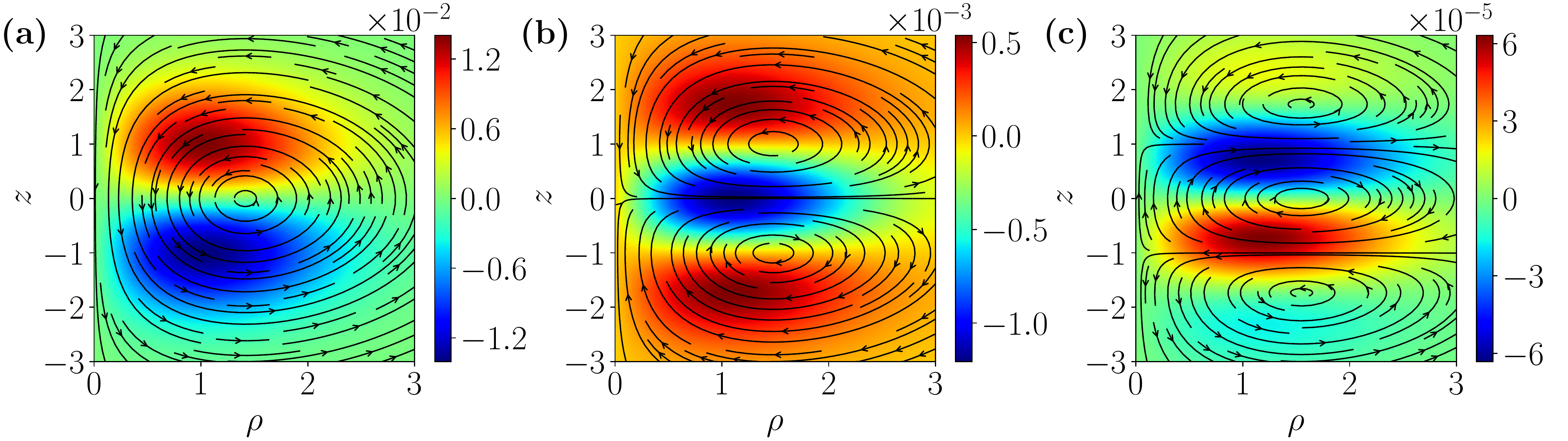}
\end{center}
\caption{The perturbative corrections to third order in $\epsilon$ of the density in cylindrical coordinates $p_{fsc}=2\pi\rho   p_{fs}=\sum_k p_{fsc}^{(k)}\epsilon^k$ and the associated current lines for the case $\alpha=10$ (corresponding to $\eta=0.2$). {\bf (a)} $p^{(1)}_{fsc}(\rho,z), \ {\bf J}^{(1)}_{fs}(\rho,z)$. {\bf (b)} $p^{(2)}_{fsc}(\rho,z),\  {\bf J}^{(2)}_{fs}(\rho,z)$. {\bf (c)} $p^{(3)}_{fsc}(\rho,z), \ {\bf J}^{(3)}_{fs}(\rho,z)$.}\label{pert}
\end{figure}

We can formally invert the Laplace transform in Eq.~(\ref{0431}) to obtain
\begin{equation}
p_{fs}(\rho,z)= \frac{\exp\left(-\frac{\rho^2}{2}\right)}{(2\pi)^{\frac{3}{2}}}\sum_{k=0}^\infty\epsilon^k C_k(\rho,\alpha)
\frac{\partial^k}{\partial z^k}\exp\left(-\frac{z^2}{2}\right) .\label{repe}
\end{equation}
We find that to order $\epsilon^3$ 
\begin{align}
p_{fs}(\rho,z)&= \frac{\exp\left(-\frac{\rho^2+z^2}{2}\right)}{(2\pi)^{\frac{3}{2}}}\left[ 1 +\epsilon z \frac{2\alpha + \rho^2}{2\alpha(\alpha+1)}+\epsilon^2(z^2-1)\frac{4\alpha^2(3+\alpha)+ 4\alpha(3+\alpha)\rho^2+(2+\alpha) \rho^4}{8\alpha^2(1+\alpha)^2(2+\alpha)}\right.\nonumber \\
&+ \left. \epsilon^3 z(z^2-3)\frac{(8 \alpha ^3+12\alpha^2 \rho^2) (\alpha^2+ 8 \alpha +19)+6 \alpha  (\alpha +3) (\alpha +4) \rho ^4+(\alpha +2) (\alpha +3) \rho ^6}{48 \alpha ^3 (\alpha +1)^3 (\alpha +2) (\alpha +3)}\right].\label{expfs}
\end{align}
One can verify that the expansion to first order in $\epsilon$ agrees with  the first order perturbation computations  of Refs.~\cite{moyses2015, mangeat2019}. It can be extended to arbitrarily higher orders. The first terms $k=1,\ 2$ and $3$ of the series  (\ref{repe}) are represented in Fig.~\ref{pert} (up to a factor $2\pi\rho$ to switch to cylindrical coordinates) for $\alpha=10$ (corresponding to a typical experimental value $\eta=0.2$). The first order correction exhibits a local maximum above $z=0$ and local minimum below. The second order term has a minimum on the axis $z=0$ and two maxima above and below. The third order result exhibits two maxima and two minima. This generation of maxima and minima is due to the generation of nodes in the functions
\begin{equation}
V_k(z)= \frac{\partial^k}{\partial z^k}\exp\left(-\frac{z^2}{2}\right).
\end{equation}
which are  related to the Hermite polynomials associated with the simple Harmonic oscillator.

\subsection{Steady state currents}

One of the key points of interest in the model here is the presence of currents which are the signature of the breaking of time reversal symmetry in the non-equilibrium steady state. The current is explicitly given by
\begin{equation}
{\bf J}_f(\rho,z) = -\left[\frac{\partial p_{fs}}{\partial \rho}(\rho,z) + \rho p_{fs}(\rho,z)\right]{\bf e}_\rho +\frac{2}{\alpha}\left[\left(\frac{\epsilon}{2\alpha}\rho^2-z \right)p_{fs}(\rho,z)-\frac{\partial p_{fs}}{\partial z}(\rho,z)\right]{\bf e}_z.
\end{equation}
To examine the current in the steady state  it is again convenient to work with the two sided Laplace transform of the current 
\begin{equation}
\tilde {\bf J}_{fs}(x,y,s) =  \int dz \exp(-s z)\ {\bf J}_{fs}(x,y,z),
\end{equation}
and from this we find
\begin{align}
\tilde{\bf J}_{fs}(x,y,s) &= -\left[\frac{\partial \tilde p_{fs}}{\partial x}(x,y,s) + x\tilde p_{fs}(x,y,s)\right]{\bf e}_x - \left[\frac{\partial \tilde p_{fs}}{\partial y}(x,y,s) + y\tilde p_{fs}(x,y,s)\right]{\bf e}_y  \nonumber \\
 &+\eta\left[\frac{\epsilon\eta}{4}(x^2+y^2) \tilde p_{fs}(x,y,z)+\frac{\partial \tilde p_{fs}}{\partial s}(x,y,s)-s \tilde p_{fs}(x,y,s)\right]{\bf e}_z.
\end{align}
It is convenient to write this in terms of the functions $A(s)$ and $B(s)$ in Eq.~(\ref{eA}) and (\ref{eB})
\begin{equation}
\tilde{\bf J}_{fs}(x,y,s)={\tilde p_{fs}(x,y,s)} \left\{ -[1-A(s)](x{\bf e}_x+ y{\bf e}_y)
 +\eta\left[\left(\frac{\epsilon\eta}{4}-\frac{A'(s)}{2}\right)(x^2+y^2) + 2\frac{B'(s)}{B(s)} \right]{\bf e}_z\right\}.\label{exjs}
\end{equation}
An interesting consequence of this result is that the current for the full system, taking into account the term $z_o$, is related to that of the system $(x,y,z_n)$ by a simple convolution in the same way as the probability density functions are related, that is to say
\begin{equation}
{\bf J}_{fs}(x,y,z) = \frac{1}{\sqrt{2\pi}}\int_0^\infty dz_n \exp\left[-\frac{1}{2}(z-z_n)^2\right]{\bf J}_{ns}(x,y,z_n).
\end{equation}
We thus see that both the steady state distribution and current of the full process $(x,y,z)$ is related to the process $(x,y,z_n)$ by a simple convolution. 

We proceed by using the representation Eq.~(\ref{repe}) to express the current as a power series in 
$\epsilon$ 
\begin{equation}
{\bf J}_{fs}(\rho,z)= \sum_{k=0}^\infty \epsilon^k {\bf J}^{(k)}_{fs}(\rho,z),
\end{equation}
where we have
\begin{equation}
{\bf J}^{(k)}_{fs}(\rho,z)= \frac{\exp\left(-\frac{\rho^2}{2}\right)}{(2\pi)^{\frac{3}{2}}}\left\{-\frac{\partial C_k}{\partial\rho}(\rho,\alpha) V_k(z) {\bf e}_\rho + \left[\frac{\rho^2}{\alpha^2}  C_{k-1}(\rho,\alpha)V_{k-1}(z)-\frac{2}{\alpha}C_{k}(\rho,\alpha)(V'_k(z)+ z V_k(z))\right]{\bf e}_z \right\}.
\end{equation}
It is obvious that $V'_k(z)= V_{k+1}(z)$ and  a straight forward exercise shows that 
\begin{equation}
z V_k(z)= -V_{k+1}(z)-kV_{k-1}(z).
\end{equation}
Using this we find
\begin{equation}
{\bf J}^{(k)}_{fs}(\rho,z)= \frac{\exp\left(-\frac{\rho^2}{2}\right)}{(2\pi)^{\frac{3}{2}}}\left\{-\frac{\partial C_k}{\partial\rho}(\rho,\alpha) V_k(z) {\bf e}_\rho + V_{k-1}(z) \left[\frac{\rho^2}{\alpha^2}  C_{k-1}(\rho,\alpha)+\frac{2k}{\alpha}C_{k}(\rho,\alpha)\right] {\bf e}_z \right\}.
\end{equation}
The divergence of the steady state current must be zero and this must hold at each order in the expansion in $\epsilon$, we thus obtain $\nabla\cdot {\bf J}^{(k)}_{fs} = 0$ for all $k$ which yields differential equations relating the polynomials $C_k(\rho,\alpha)$ which can be used to generate them iteratively starting at $C_0(\rho,\alpha)=1$. 

The first three orders in perturbation theory for the current then give
\begin{align}
{\bf J}^{(1)}_{fs}(\rho,z) &= \frac{\exp\left(-\frac{\rho^2+z^2}{2} \right)}{(2\pi)^\frac{3}{2}}\left[-\frac{\rho  z}{\alpha  (\alpha +1)} {\bf e}_\rho + \frac{\rho ^2-2}{\alpha  (\alpha +1)} {\bf e}_z\right]\\
{\bf J}^{(2)}_{fs}(\rho,z) &= \frac{\exp\left(-\frac{\rho^2+z^2}{2} \right)}{(2\pi)^\frac{3}{2}}\left[- \frac{\rho  \left(z^2-1\right) \left[(\alpha +2) \rho ^2+2 \alpha  (\alpha +3)\right]}{2
   \alpha ^2 (\alpha +1)^2 (\alpha +2)} {\bf e}_\rho\right. \nonumber\\
   &+\left. \frac{z \left[(\alpha +2) \rho ^4 + 2 \left(\alpha ^2+\alpha -4\right) \rho ^2 -4 \alpha 
   (\alpha +3)\right]}{2 \alpha ^2 (\alpha +1)^2 (\alpha +2)}{\bf e}_z \right]\\
{\bf J}^{(3)}_{fs}(\rho,z) &= \frac{\exp\left(-\frac{\rho^2+z^2}{2} \right)}{(2\pi)^\frac{3}{2}}\left[- \frac{\rho  z \left(z^2-3\right) \left[(\alpha +2) (\alpha +3) \rho ^4+4 \alpha  (\alpha +3) (\alpha +4) \rho ^2 + 4 \alpha ^2 (\alpha^2+ 8 \alpha +19)\right]}{8 \alpha ^3
   (\alpha +1)^3 (\alpha +2) (\alpha +3)} {\bf e}_\rho\right. \nonumber \\
&+\left. \frac{\left(z^2-1\right) \left[(\alpha +2) (\alpha
   +3) \rho ^6+2 (\alpha +3) ( 2\alpha^2 +5\alpha-6) \rho ^4+4 \alpha  (\alpha^3 +4\alpha^2-9\alpha-48) \rho ^2 -8 \alpha ^2 (\alpha^2+ 8\alpha +19)\right]}{8 \alpha ^3 (\alpha +1)^3 (\alpha +2) (\alpha +3)}{\bf e}_z\right]
\end{align}
One can again verify that the expansion to first order in $\epsilon$ agrees with  the first order perturbation computations  of Refs.~\cite{moyses2015, mangeat2019}. 

The first three perturbative corrections to the steady state current are shown in Fig.~\ref{pert} for the case $\alpha=10$ superimposed on the corresponding corrections to the steady state probability density function of $\rho$ and $z$. We see that the first order correction possesses and single vortex, while the second and third order corrections have two and three vortices respectively. A thorough numerical investigation however rules out the existence of more than one vortex and when fully summed the vortices seen in the individual terms of the perturbation almost certainly vanish. Indeed it seems physically unlikely that a tracer current circulates around more that one central vortex. When $\rho$ is large, particles are pushed upwards by the non-conservative force and when it becomes small enough the harmonic restoring term pulls them down. 

\subsection{The circulation of the current}

Of particular interest is the circulation of the non-equilibrium current defined by
\begin{equation}
\Omega_{fs}= \frac{1}{2\pi}\int d{\bf x}\ \nabla\times {\bf J}_{fs}({\bf x})\cdot{\bf e}_\phi
\end{equation}
where  ${\bf e}_\phi= (-y {\bf e}_x+ x {\bf e}_y)/\rho$ is the basis vector of the polar angle $\phi$, and by symmetry the local current circulates around this direction. Here is we use Eq.~(\ref{exjs}) for steady state current written in polar coordinates 
\begin{equation}
\tilde{\bf J}_{fs}(\rho,s)={\rho\tilde p_{fs}(\rho,s)} \left\{ -[1-A(s)]{\bf e}_\rho
 +\eta\left[\left(\frac{\epsilon\eta}{4}-\frac{A'(s)}{2}\right)\rho^2 + 2\frac{B'(s)}{B(s)} \right]{\bf e}_z\right\},\label{exjsc}
\end{equation}
which yields
\begin{equation}
\Omega_{fs}= \int dz \rho d\rho\left[ \frac{\partial }{\partial z}J_{fs,\rho} - \frac{\partial }{\partial \rho}J_{fs,z}\right] =  \int d\rho \tilde J_{fs,z}(\rho,0).
\end{equation}
In the above we have used that the first term in the integrand integrates directly giving zero and the second term has been integrated by parts. Using the explicit form of ${\bf J}_{fs}$ in Eq.~(\ref{exjsc}) together with Eqs. (\ref{EqA}), (\ref{EqB}), (\ref{fulln}), we obtain
\begin{equation}
\Omega_{fs}= -\frac{\eta^2\epsilon}{4\sqrt{2\pi}(\eta+2)}.
\end{equation}
Interestingly this  is in agreement, up to a change in sign as we have reversed the direction of propagation of the laser here, with the first order perturbation calculation of the circulation given in Ref.~\cite{moyses2015}. The first order perturbative result of  Ref.~\cite{moyses2015} thus turns out to be exact for all values of $\epsilon$.

\section{Conclusions} 
\label{secDiscussion}
In this paper we have analyzed the simplest model used for optical traps that takes into account non-conservative forces generated by the laser interaction with the trapped particle's dipole moment. Non-conservative forces occur frequently in physics, notably due to the magnetic components of electromagnetic fields, however analytical descriptions of the associated steady states are very rare and no general theory exists. We were able to find exact results for the Laplace transform with respect to the coordinate  $z$ (the direction of laser propagation) of the full non-equilibrium steady state probability distribution. In the limit $\varepsilon=\infty$, $z$ can be replaced by the non-conservative part $z_n$. We have also computed the Laplace transform of the marginal distribution of the nonconservative part of the displacement in $z$, this can be formally inverted and the full distribution found.   
The full distribution of $z$ is then simply related to this latter distribution via a convolution with a Gaussian. Interestingly the same convolution relationship is found to hold for the currents. The Laplace transform representation of the full three dimensional probability distribution can be used to develop a systematic perturbation theory of the full steady state probability distribution as a series in the magnitude of the non-equilibrium force, which here is  denoted by $\epsilon$. In this paper  we have given explicit results to ${\cal O}(\epsilon^3)$, extending the first order results of \cite{moyses2015,mangeat2019}, but arbitrary higher orders can be deduced from our formalism.  Using our results, we were also able to compute the circulation of the current in the steady states, interestingly this exact results agrees with that of first order perturbation theory found in \cite{moyses2015}.  

In \cite{mangeat2019} the same model but with under damped dynamics was studied. The presence of non-conservative forces means that the steady state depends on the friction  in the system and the velocity and spatial degrees of freedom are not independent as is the case with the equilibrium Gibbs-Boltzmann distribution. The results of \cite{mangeat2019} were perturbative and it would be interesting to see if the method used here could be used to exactly analyze the underdamped model. 

\section{Acknowledgements}

M.M. was partially financially supported by the German Research Foundation (DFG) within the Collaborative Research Center SFB 1027.


\begin{thebibliography}{46}

\bibitem{ashkin1970} A. Ashkin, {\itshape Acceleration and Trapping of Particles by Radiation Pressure}, Phys. Rev. Lett. {\bfseries 24}, 156-159 (1970).

\bibitem{ashkin2006} A. Ashkin, {\itshape Summary of the First Decade's Work on Optical Trapping and Manipulation of Particles} in {Optical Trapping and Manipulation of Neutral Particles Using Lasers}, World Scientific (2006).

\bibitem{ashkin1987} A. Ashkin and J. M. Dziedzic, {\itshape Optical trapping and manipulation of viruses and bacteria}, Science {\bfseries 235}, 1517-1520 (1987).

\bibitem{grier1997} D. G. Grier, {\itshape Optical tweezers in colloid and interface science}, Current Opinion in Colloid \& Interface Science {\bfseries 2}, 264-270 (1997).

\bibitem{lukic2007} B. Lukic, S. Jeney, Z. Sviben, A. J. Kulik, E.-L. Florin, and L. Forro, {\itshape Motion of a colloidal particle in an optical trap}, Phys. Rev. E {\bfseries 76}, 011112 (2007).

\bibitem{jop2009} P. Jop, J. R. Gomez-Solano, A. Petrosyan, and S. Ciliberto, {\itshape Experimental study of out-of-equilibrium fluctuations in a colloidal suspension of Laponite using optical traps}, J. Stat. Mech. {\bfseries 2009}, P04012 (2009).

\bibitem{dienerowitz2008} M. Dienerowitz, M. Mazilu, and K. Dholakia, {\itshape Optical manipulation of nanoparticles: a review}, JNP {\bfseries 2}, 021875 (2008).

\bibitem{gieseler2012} J. Gieseler, B. Deutsch, R. Quidant, and L. Novotny, {\itshape Subkelvin Parametric Feedback Cooling of a Laser-Trapped Nanoparticle}, Phys. Rev. Lett. {\bfseries 109}, 103603 (2012).

\bibitem{li2013} T. Li, M. G. Raizen, {\itshape Brownian motion at short time scales}, Annalen der Physik {\bfseries 525}, 281 (2013).

\bibitem{bateman2014} J. Bateman, S. Nimmrichter, K. Hornberger, and H. Ulbricht, {\itshape Near-field interferometry of a free-falling nanoparticle from a point-like source}, Nat. Comm. {\bfseries 5}, 4788 (2014).

\bibitem{lehmuskero2015} A. Lehmuskero, P. Johansson, H. Rubinsztein-Dunlop, L. Tong, and M. K\"all, {\itshape Laser Trapping of Colloidal Metal Nanoparticles}, ACS Nano {\bfseries 9}, 3453-3469 (2015).

\bibitem{wildermuth2004} S. Wildermuth, P. Kr\"uger, C. Becker, M. Brajdic, S. Haupt, A. Kasper, R. Folman, and J. Schmiedmayer, {\itshape Optimized magneto-optical trap for experiments with ultracold atoms near surfaces}, Phys. Rev. A {\bfseries 69}, 030901 (2004).

\bibitem{chu1991} S. Chu, {\itshape Laser Manipulation of Atoms and Particles}, Science {\bfseries 253}, 861-866 (1991).

\bibitem{wang1997} M. D. Wang, H. Yin, R. Landick, J. Gelles, and S. M. Block, {\itshape Stretching DNA with optical tweezers}, Biophysical J. {\bfseries 72}, 1335-1346 (1997).

\bibitem{bustamante2000} C. Bustamante, S. B. Smith, J. Liphardt, and D. Smith, {\itshape Single-molecule studies of DNA mechanics}, Current Opinion in Structural Biology {\bfseries 3}, 279-285 (2000).

\bibitem{huguet2010} J. M. Huguet, C. Bizarro, N. Forns, S. B. Smith, C. Bustamante, and F. Ritort, {\itshape Single-molecule derivation of salt dependent base-pair free energies in DNA}, PNAS {\bfseries 107}, 15431-15436 (2010).

\bibitem{wen2007} J.-D. Wen, M. Manosas, P. T. X. Li, S. B. Smith, C. Bustamante, F. Ritort, and I. Tinoco , {\itshape Force Unfolding Kinetics of RNA Using Optical Tweezers. I. Effects of Experimental Variables on Measured Results}, Biophysical J. {\bfseries 92}, 2996-3009 (2007).

\bibitem{ritort2006} F. Ritort, S. Mihardja, S. B. Smith, and C. Bustamante, {\itshape Condensation Transition in DNA-Polyaminoamide Dendrimer Fibers Studied Using Optical Tweezers}, Phys. Rev. Lett. {\bfseries 96}, 118301 (2006).

\bibitem{moore2014} D. C. Moore, A. D. Rider, and G. Gratta, {\itshape Search for Millicharged Particles Using Optically Levitated Microspheres}, Phys. Rev. Lett. {\bfseries 113}, 251801 (2014).

\bibitem{ranjit2016} G. Ranjit, M. Cunningham, K. Casey, and A. A. Geraci, {\itshape Zeptonewton force sensing with nanospheres in an optical lattice}, Phys. Rev. A {\bfseries 93}, 053801 (2016).

\bibitem{liu2017} J. Liu and K.-D. Zhu, {\itshape Nanogravity gradiometer based on a sharp optical nonlinearity in a levitated particle optomechanical system}, Phys. Rev. D {\bfseries 95}, 044014 (2017).

\bibitem{arvanitaki2013} A. Arvanitaki and A. A. Geraci, {\itshape Detecting High-Frequency Gravitational Waves with Optically Levitated Sensors}, Phys. Rev. Lett. {\bfseries 110}, 071105 (2013).

\bibitem{rider2016} A. D. Rider, D. C. Moore, C. P. Blakemore, M. Louis, M. Lu, and G. Gratta, {\itshape Search for Screened Interactions Associated with Dark Energy below the 100 µm Length Scale}, Phys. Rev. Lett. {\bfseries 117}, 101101 (2016).

\bibitem{collin2005} D. Collin, F. Ritort, C. Jarzynski, S. B. Smith, I. Tinoco Jr, and C. Bustamante, {\itshape Verification of the Crooks fluctuation theorem and recovery of RNA folding free energies}, Nature {\bfseries 437}, 7056 (2005).

\bibitem{carberry2004} D. M. Carberry, J. C. Reid, G. M. Wang, E. M. Sevick, D. J. Searles, and D. J. Evans, {\itshape Fluctuations and Irreversibility: An Experimental Demonstration of a Second-Law-Like Theorem Using a Colloidal Particle Held in an Optical Trap}, Phys. Rev. Lett. {\bfseries 92}, 140601 (2004).

\bibitem{carberry2007} D. M. Carberry, M. A. B Baker, G. M. Wang, E. M. Sevick, and D. J. Evans, {\itshape An optical trap experiment to demonstrate fluctuation theorems in viscoelastic media}, J. Opt. A: Pure Appl. Opt. {\bfseries 9}, 5204 (2007).

\bibitem{berut2016} A. B\'erut, A. Imparato, A. Petrosyan, and S. Ciliberto, {\itshape Stationary and Transient Fluctuation Theorems for Effective Heat Fluxes between Hydrodynamically Coupled Particles in Optical Traps}, Phys. Rev. Lett. {\bfseries 116}, 068301 (2016).

\bibitem{bassi2013} A. Bassi, K. Lochan, S. Satin, T. P. Singh, and H. Ulbricht, {\itshape Models of wave-function collapse, underlying theories, and experimental tests}, Rev. Mod. Phys. {\bfseries 85}, 471-527 (2013).

\bibitem{wu2009} P. Wu, R. Huang, C. Tischer, A. Jonas, and E.-L. Florin, {\itshape Direct Measurement of the Nonconservative Force Field Generated by Optical Tweezers}, Phys. Rev. Lett. {\bfseries 103}, 108101 (2009).

\bibitem{roichmann2008} Y. Roichmann, B. Sun, A. Stolarski, and D. G. Grier, {\itshape Influence of Nonconservative Optical Forces on the Dynamics of Optically Trapped Colloidal Spheres: The Fountain of Probability}, Phys. Rev. Lett. {\bfseries 101}, 128301 (2008).

\bibitem{sun2009} B. Sun, J. Lin, E. Darby, A. Y. Grosberg, and D. G. Grier, {\itshape Brownian vortexes}, Phys. Rev. E {\bfseries 80}, 010401 (2009).

\bibitem{sun2010} B. Sun, D. G. Grier, and A. Y. Grosberg, {\itshape Minimal model for Brownian vortexes}, Phys. Rev. E {\bfseries 82}, 021123 (2010).

\bibitem{simpson2010} S. H. Simpson and S. Hanna, {\itshape First-order nonconservative motion of optically trapped nonspherical particles}, Phys. Rev. E {\bfseries 82}, 031141 (2010).

\bibitem{moyses2015} H. W. Moyses, R. O. Bauer, A. Y. Grosberg, and D. G. Grier, {\itshape Perturbative theory for Brownian vortexes}, Phys. Rev. E {\bfseries 91}, 062144 (2015).

\bibitem{demessieres2011} M. de Messieres, N. A. Denesyuk, and A. La Porta, {\itshape Noise associated with nonconservative forces in optical traps}, Phys. Rev. E {\bfseries 84}, 031108 (2011).

\bibitem{amarouchene2019} Y. Amarouchene, M. Mangeat, B. V. Montes, L. Ondic, T. Gu\'erin, D. S. Dean, and Y. Louyer, {\itshape Nonequilibrium Dynamics Induced by Scattering Forces for Optically Trapped Nanoparticles in Strongly Inertial Regimes}, Phys. Rev. Lett. {\bfseries 122}, 183901 (2019).

\bibitem{mangeat2019} M. Mangeat, Y. Amarouchene, Y. Louyer, T. Gu\'erin, and D. S. Dean, {\itshape Role of nonconservative scattering forces and damping on Brownian particles in optical traps}, Phys. Rev. E {\bfseries 99}, 052107 (2019).

\bibitem{liver2020} T. B. Liverpool, {\itshape Steady-state distributions and nonsteady dynamics in nonequilibrium systems}, Phys. Rev. E {\bfseries101}, 042107 (2021).

\bibitem{gieseler2013} J. Gieseler, L. Novotny, and R. Quidant, {\itshape Thermal nonlinearities in a nanomechanical oscillator}, Nat. Phys. {\bfseries 9}, 806 (2013).

\bibitem{boyer2011} D. Boyer and D. S. Dean, {\itshape On the distribution of estimators of diffusion constants for Brownian motion}, J. Phys. A: Math. Theor. {\bfseries 44}, 335003 (2011).

\bibitem{abram} M. Abramowitz and I. A. Stegun, {\itshape Handbook of Mathematical Tables}, (Dover, New York, 1965).
\bibitem{colo2017}I. Colombaro, A. Giusti and F. Mainardi, {\itshape A class of linear viscoelastic models based on Bessel functions}, Meccanica {\bfseries 52}, 825 (2017).

\bibitem{kent1978} J. T. Kent, {\itshape Some probabilistic properties of Bessel functions}, Ann. Probab. {\bfseries 6},760 (1978).

\bibitem{geto1979} R. K. Getoor and M. J. Sharpe, {\itshape Excursions of Brownian motion and Bessel processes}, Z. Wahr. Ver. Gebiete {\bfseries 47}, 83 (1979).

\end{thebibliography}
\end{document}